*Interactionalism:*

# *Re-Designing Higher Learning for the Large Language Agent Era*

Mihnea Moldoveanu* and George Siemens**

*University of Toronto; mihnea.moldoveanu@rotman.utoronto.ca

**Matter and Space, Inc.and the University of South Australia;  gsiemens@matterandspace.com

## *Abstract*

We introduce *Interactionalism* as a new, internally coherent set of guiding principles and heuristics for the design and architecture of learning now available due to Generative AI (GenAI). It is focused on explicit pedagogical moves, gestures and maneuvers that focus on the dialogical and dynamics of learning together with GenAI agents. The value of this approach is in anticipation of growing reliance on, and integration with, AI as a co-thinker with humans, and an important factor in the production function of 'skills' and 'knowledge' across many domains. We introduce and articulate *interactional intelligence* as a net new skill set that is increasingly important when core cognitive tasks are automatable and augmentable by GenAI functions. We break down these skills into meta-cognitive and meta-emotional components and show how working with Large Language Model (LLM)-based agents – or, Large Language Agents (LLA's) - can help learners develop them. Interactionalism is advanced as a blueprint for the *practice of GenAI-first learning* - and interactional intelligence is presented as a net new skill set that courses, programs and institutions should aspire to help learners develop.

## *Introduction*

All educational fields have been significantly impacted by the broad availability and use of Large Language Models and agentic infrastructures based on them. Educational systems have made local adjustments to the new learning environment, introducing both measures to mitigate the occurrence and adverse consequences of cognitive offloading and algiarism, and the opportunities for feedback and virtual teaching assistance LLM's provide. However, there is an unfulfilled need for a paradigmatic shift in the way teaching and learning 'happen', which can serve as a guiding vision for the re-design of existing programs of study and the design of *de novo,* AI-first programs. We introduce *Interactionalism*[1] as a framework and set of guiding principles for learning in the (GenAI) era. GenAI – and in particular dialogical agents based on LLM's – enable and facilitate learning that is personalized, socialized and *conversational* in nature – in sharp contrast to linear, standardized approaches largely extant in higher education that are based on monological approaches to content presentation and skill assessment and one-to-many ('broadcasting') approaches to knowledge dissemination. This new learning approach matches and tracks the ways in which the nature of human work - tasks, skills and roles – has shifted over the past two decades, towards greater levels of interactivity and collaboration.

An interactionalist approach to learning closely matches the current landscape of human labor markets [Moldoveanu, 2024]. In this new regime, *know*-how remains privileged relative to know-what (as 'information is one click away') - as it has been for the past two decades [Siemens, 2005]. But learning *together* with AI – and learning to *think* and *act* together with AI and with other humans in ways moderated and facilitated by AI agents – are *net new* capabilities and corresponding skills we can aim to have learners develop by *processes that themselves become models* of the desired skills. We articulate *interactional intelligence* as a skill set, break it down into core components, establish links to meta-cognitive and meta-emotional[2] skills and show how Large Language Agents ('LLA's') – acting as dialogical agents across learning scenarios and powered by LLM's that like flexible operating systems - can be used to help learners develop it. We outline ways in which schools and universities can implement architectural re-designs of their entire set of learning activities - using GenAI that exists today – to embody the interactionalist approach to learning. A key component of an interactionalist approach to learning is a recognition and sharp definition of dialogical LLA's that capture with fidelity, precision and nuance the conversational and interactional structures of teaching and learning and constitute the 'fabric of learning'. The design of such agents – as well as their use – both relies on interactional intelligence and helps humans develop key components of it. In contradistinction to approaches that aim to *lower* the cognitive and meta-cognitive load that interacting with LLM's and designing LLA's creates [Tankelevitch et al, 2024], we focus on the opportunity to use the new dialogical and interactionalist landscape of learning as a way of helping learners develop and expand their 'meta-human' skill set.

[1] There are several senses in which *interaction**ism*** has been used before, notably in developmental psychology, the theory of mind-body interactions and the literature dealing with **interactions** between observer and phenomenon in quantum mechanics. We have called our approach *interaction**alism*** to distinguish it from *interactionism* while retaining the *interactional* form of the tasks and skills it comprises.

Interactionalism builds on growing interest in human-computer interaction that focuses not only on the interplay and coordination of the cognitive outputs that are passed back and forth between humans and machines (AI) across a fixed boundary, but also makes understanding machine behavior a primary intent (Rahwan et al. , 2019,, Taylor & Taylor, 2021) and allow the boundary to evolve adaptively to the tasks at hand. The evaluation of how machines 'think' and operate are foundational to interactionalism as this determines the types of engagement (i.e. pedagogical maneuvers, in an educational) that LLA's can engage in. We address these specific approaches below through the lens of GenAI's ability to create novel outputs in real time – a significant new affordance not evident in today's textbook and lecture-based learning models.

[2] Whereas metacognitive skills have received detailed treatments and the phrase has become a term of art, 'meta-emotional' skills may be rather less familiar. We mean them to represent a reliably exercised proclivity and ability to not only comprehend and articulate the feelings , sentiments and visceral qualia evoked in a person by a possible interaction or dialogue with another, but also to proactively design and shape interventions that produce desired emotional dynamics. It is a skill commonly associated with *dramaturges,* playwrights and screenwriters, whose work has much in common with that of an LLA designer.

*GenAI in Education: The Challenge and Opportunity Set*

The challenge GenAI poses to the traditional models of education is clear: at a time when learners can use GenAI tools to replicate 'good enough' answers to the questions, assignments, and quizzes that form the canon of existing learning assessment, which are only weakly, if at all, identifiable as having been AI-generated, the integrity of individual assessment is imperilled. Thereby, the *signaling value of certifications of learning* bestowed upon learners that have access to such tools and are evaluated by standard models of education is reduced [Moldoveanu, 2024] and therefore the labor market value of the experiences that comprise an educational system is also diminished. The 'GenAI' challenge to assessment integrity is real, in the context of the standard models of education.

Independently of the assessment integrity crisis, there is a challenge in designing the interactions between GenAI and humans that occur in the learning process that enhance and promote learner agency, engagement and curiosity and . A common 'use case has been to have GenAI serve as a teaching assistant or tutor that guides and directs learners. This application is useful but its successful deployment requires the development of an effective *dialogical grammar* that structures the ways in which an LLA directs learners through questioning, answer provisioning, and stimulating more precise and sophisticated responses. Some studies of the uses of GenAI in education show that learners tend to rely on AI for learning, but don't learn from it (Darvishi, et al, 2024). There is, thus, ample room for both experimentation and improvement in the patterning of dialogical use cases and a grammar of learning-oriented dialogue (or, a dialogical grammar) that promotes and enhances learning.

Understanding the challenges and the opportunities GenAI presents can be advanced by reframing the key assumptions and commitments that standard models of education share:

- To skill or competence as being individual in nature, in the sense that they are exercised, embodied and learned individually, assessed on an individual, 'one-shot' basis (one question- one answer) and designed to be deployed on an individual basis;

- To know-how being associated with the completion of an individual task at a certain level of quality (accuracy, validity, coherence, verifiability) and reliability in a certain, pre-set period of time;
- To an incontrovertible base of 'know-what' comprising facts, rules, associations, laws, norms that undergird the individual performance of the tasks comprising the learner's know-how.

These commitments together make current educational models vulnerable to *assessment integrity* challenges from learners equipped with access to GenAI tools, which, by design, are competent at reproducing 'probably, approximately correct', - or, normative - answers to canonical questions. However, these commitments are also precisely the ones we now need to re-visit and revise, for reasons related to the ways human work has come to be organized during the past thirty years [Johnson, Manyika and Yee, 2007; Deming, 2017; Garicano, 2000]:

- Skill and competence are largely exercised interactively and dialogically, through assertions, questions, queries, challenges, answers and responses that are directed, directive and embedded – in a context, with a pretext and a context, and for a dialogical and conversational purpose [Moldoveanu, 2024].

- The tasks that are yardsticks for competence are themselves interactive and dialogical in nature, rather than carried out 'by oneself', in a way that is uninterrupted by interjections, questions, challenges and adaptive, iterative re-framings. They often entail or require dialogue - in synchronous or asynchronous form - and their criteria for successful closure are dialogical ('agreement', 'clarity on areas of disagreement') rather than monological ('optimality', 'consistency', etc) in nature [Moldoveanu and Narayandas, 2022];.

- The base of know-what that grounds the exercise of competent know-how is almost never beyond doubt or incontrovertible. Rather, it is subject to interactive and collective processes, often with peers, and framing, articulation, inference and deliberation that are social and inter-personal in nature. 'Facts' are footnoted by epistemological metadata no one who wants to count on them can ignore because they qualify validity, accuracy and reliability, and which can only be brought into relief by queries, questions and challenges [Moldoveanu and Martin, 2008];

- The growing recognition that many human cognitive tasks can be carried out with high reliability and precision (i.e. ‘automated’) by AI agents creates a labor market environment which pressures an educational system to focus on developing quintessentially human skills [Moldoveanu, 2024] and specifically to address epistemological (‘how do you know X to be true/valid/explanatory’) and ontological dimensions of humans’ ways of being in the world and being with other humans [Brey, 2005].

The GenAI challenge to the assessment integrity of current models is a timely signal of a structural change that is now both due and feasible. That change is one that takes us to *dialogical learning*: a model of learning that is interactive, conversational and non-linear, by distinction from standard monological and linear practices. In the standard model, a course is a linear object: a set of reading or viewing materials (‘lectures’ and ‘readings’) a set of assignments and quizzes ordered to succeed each one of the ‘intake’ episodes, and an assessment instrument (‘exam’ or ‘test’) that tests for a combination of know-how and know-what the learner is to have acquired. Feedback in this model is scant, generic and delayed relative to the learner’s performances [Moldoveanu and Djikic, 2017].

By contrast, the dialogical model lays out the set of core competencies or skills the learner needs to have demonstrated to have ‘passed’ the course and allows the learner the agency to engage directly with a dialogical agent that functions as a tutor, teaching assistant, evaluator, guide and mentor. The learner and the dialogical LLA are together responsible for structuring the material of the course, and for creating and making the specific, in-the-moment choices (‘read further or take a quiz’, ‘do an assignment or have an open-ended discussion’,’engage in a debate or simulate an interview on…’, ‘play out a simulation of… or create a new app that …’etc). Some 40 years ago, Benjamin Bloom [Bloom, 1984] showed that one on one tutoring of students leads to a ‘2 sigma effect’, in that students that learn through dialogical learning outperform 98% of their peers that learn through standard methods – an effect equivalent to that of a C-level student becoming an A-level student. The challenge has always been that one-on-one attention to the learner is too expensive, in time and resources. GenAI allows us to address the ‘learner intimacy challenge’ head on and design the ‘always on’ tutor that walks the learner, step by step, through the process of skill acquisition.

Such a learning process is fundamentally a shift from linear to network approaches. It mirrors, in its structure and function, changes in human work in organizations, as the tasks for which learners need to be prepared is rapidly changing as well as a result of the use of GenAI in the workplace [Dell Aqua et al, 2023; Mollick, 2024]. The base of skills most sought after by the organizations that recruit and employ graduates has shifted towards the skills required to design and take part in *interactions* [Johnson, Manyika and Yee, 2007; Deming, 2015] and those skills that are predominantly *communicative* and *dialogical* in nature [Moldoveanu, 2024]. We have, for some time, been moving from an era of skills that are individually learned and exercised to an era in which they are interactionally and dialogically exercised, and whose acquisition will greatly benefit from being *interactionally and dialogically learned* as well. GenAI – and particularly the dialogical learning and interaction afforded by LLA’s – present an opportunity to re-design and re-engineer the fabric of learning activities so that they match the new base of skills and competencies learners even now are desired – and soon required – to have developed.

This transition to more active and engaged learning calls into question more than just the mechanics of specific, local instructor-learner interactions. Our collective language for referring to the elements of a learning experience still carries within it the ontology of the classical university experience: courses, lectures, and readings. With a set of dialogical agents as the primary integrator of learners, instructors and content, many incumbent elements of traditional learning need to be discarded or transformed as we focus capability of an individual learner to engage with GenAI in ways that enable and facilitate the learning of GenAI-relevant skills needed in the workplace.

*Interactional Intelligence: Dialogical, Multi-Modal, Meta-Human*

Relevant prior work has recognized the importance of artificial cognition and of its working alongside human cognition [Siemens, et al., 2022] in different task and knowledge domains, and called for ‘interfacial’ research that seeks to track the comparative advantage one form of cognition has over another in various predicaments.

Cognitive approaches to skill development have relied heavily on the mind as a computational device that mimics, in allocation of memory, computation, and input-output functions, a digital computer [Johnson Laird, 1980; Gigerenzer, 2001]. In the cognitivist tradition, 'working with' computers entails drawing the right boundaries between human and artificial cognition [Siemens et al, 2022] in ways that heed the comparative advantages of humans and machines in various tasks. Just as the calculator made it possible for humans to offload large calculations and the word processor enabled the augmentation of a person's 'working memory', just so the highly flexible, self-refining algorithms enabled by deep learning enabled humans to offload tasks such as statistical inference and data classification, and to move the processing that previously happened 'offline' to an 'online', always-on, connected and fluid environment [Siemens, 2005]. Individual, cognitive skills remained privileged in educational systems, over the interactional and communicative skills that have been steadily growing in value. The capacity of GenAI technologies – and particularly LLM's currently exemplified by GPT 4, Claude 3, Llama, and Gemini – to mimic and generate dialogical interactions sensitive and responsive to conversational implicature (the ways in which spoken words refer to an unspoken shared background context), burden of proof dynamics (*who* needs to justify or explain oneself at this point in the conversation and why?) and speaker role and background as well as to the purpose and substance of each act of speaking and writing is a *quantitative advance that enables a qualitative change* in what is possible and desirable in a learning environment. That is because much of the cognitive work that humans used to have to do to be able to interact productively with machines now becomes predominantly *interactive* and often meta-cognitive work [Tankelevitch et al, 2024]. The GenAI-enabled worker 'works alongside' artificially sapient – and even artificially sentient - agents to dialogically co-create artifacts, to design cognitive tasks and task architectures, as well as tests and test beds for the output of machines for which a human will need, at the end of the workday, to take *human* responsibility [Moldoveanu, 2024; Mollick, 2024; Kosslyn, 2024]. The 'cognitive' dimension of 'knowledge work' needs to be urgently augmented by an epistemological dimension ('how do we know? what counts as knowing in this context? what counts as a valid and reliable explanation? a justification? a clarification? a model? an inference'?)

and an ontological dimension ('what is real? how do words pick out objects and events in ways that count as reliable 'reference'?) which inform the questions that will orient the re-design of work: 'What is human agency? What part of it can be assigned to machines?' 'What is machine agency?')

*Re-writing the Canon.*

This interactional way of being can be more or less skilled and the skills associated with interactions can be understood, quantified, measured and developed in ways similar to those in which we have helped learners develop cognitive, technical and algorithmic skills. Doing so requires we conceptualize a *new kind of intelligence* that is interactive and dialogical – rather than individualistic and monological – in nature:

- 'Reading' materials – a quintessentially 'individual' activity – is now carried out as a dialogue between the 'reader' and a dialogical agent capable of summarizing, extracting, paraphrasing the source material – which includes audio, video and text files – and thus requires 'dialogical, adaptive inquiry' skills, as opposed to the more or less passive processing of semantic and syntactic information. The dialogical agent may itself be *designed* by the reader to answer specific kinds of queries or produce abstractive summaries that serve particular communicative functions (like: explanation, justification, debate, deliberation) and answer to specific challenges (to validity, reliability, coverage, backing);

- 'Analysis' – which includes parsing, classifying, optimizing, and making inferences that reach beyond a body of information given – now becomes a process by which patterned queries, questions and challenges can guide a large Language Agent to make sequential, recursively refinable inferences that are themselves guided by previous interactions in the same dialogue. Alternatively, the design of such an LLA itself becomes a part of the learner's 'production function';

- 'Writing' – perhaps *the* most prevalently individual task, to which one devotes a significant amount of 'thinking pre-work' – now emerges as an interactive process of prompting, outlining, developing, elaborating, editing, culling, curating and shaping in collaboration with an AI agent that acts as a thinking, writing and editing partner, and as an interlocutor in developing and shaping the substantive foundations and inferential chains of arguments;

- Coding, traditionally an ‘individual sport’, has already become a ‘team sport’ – both via ‘paired’ or ‘collective’ programming – and via coding co-pilot assistants that offer personalized, contextualized guidance on almost every task in the software development workflow, ranging from ‘algorithm design’ to data structuration to syntactical assistance, testing and de-bugging;

- Modeling and model-building – whether for predictive or explanatory purposes and using causal, teleological or functional logics of inference - in the domains of both the natural and the social sciences is set to become an exercise in the iterative and interactive sharpening of framing questions, articulation of axiomatic assumptions, plausible inferences and testable hypotheses, along with minimal conditions for maximally informative empirical tests subject to the validity and relevance standards of a field of inquiry and practice.

The scenarios and ‘use cases’ in which writing, coding, analysis, reading and most of the ‘individualistically exercised’ skills people possess show up in the workplace are so different from the ways in which they have typically been learned and are for the most part still taught that referring to them by their accepted names can lead to confusion. For instance, ‘writing’ reliably evokes imagery of an individual sitting in front of a screen or an open notebook. These traditionally foundational skills should, rather, be augmented or replaced by their *interactively* exercised counterparts:

> *Reading→ Interactive processing of text:* conversational agents can be prompted and contextualized to produce targeted summaries of reports, videos and audio files, to process their text in ways that return responses to targeted questions and queries, to synthesize the implications of the assumptions, inferences and arguments of a piece of text to another body of textual knowledge. Conversational AI has shown great promise in helping users read and comprehend complex texts interactively. Recent large language model (LLM) agents can summarize documents or answer questions by drawing on provided texts, essentially serving as intelligent reading companions. For instance, MuDoC is a multimodal conversational agent designed as an “intelligent textbook” assistant (Taneja & Goel, 2025). MuDoC processes long textbook chapters, including both textual content and figures, and generates grounded responses that include relevant text summaries and embedded diagrams. A key feature of this system is its emphasis on trustworthiness: the interface links every answer back to the source material, allowing users to instantly navigate to the original passage or figure for verification. This instills confidence that the agent's summaries and explanations truly reflect the input text, addressing the common LLM issue of hallucinating details not in the source.

Another example involves evaluating ChatGPT's ability to answer questions based on documents in an information-seeking dialogue setting. This comparison between a vanilla ChatGPT model and a retrieval-augmented approach shows how document grounding improves answer fidelity (Braunschweiler, Doddipatla, Keizer & Stoyanchev, 2023). Human evaluations revealed that ChatGPT's answers were often more comprehensive and preferred over human-written responses, though they sometimes included information not present in the documents. This underscores the importance of grounding strategies, such as those used in MuDoC, to prevent hallucinations and provide source-verifiable content. Beyond factual Q&A and summaries, conversational agents can also assist readers in critical analysis of texts. By prompting an agent with pointed questions (e.g., "What assumptions underlie this argument?" or "How do these findings compare to earlier claims?"), learners can engage in a dialogue that teases out the implications and inferences within a document. Advanced agents are capable of discussing an author's reasoning, identifying potential biases or assumptions, and synthesizing arguments from multiple sections. This turns reading into an interactive process: the student and AI can co-explore the material, with the AI clarifying points and the student probing deeper. Such interactive processing mirrors a Socratic tutor, enhancing comprehension and critical engagement.

*Writing →Interactive creation and production of text and other media*: conversational agents can be patterned to create outlines and summaries that adequately respond to context, to refine and develop arguments, to address counter-arguments, to raise questions and challenges that would naturally occur to members of the target audience, to expand arguments via examples, examples, counterexamples and alternative narratives, and to deconstruct arguments to core assumptions and the inferences that emanate from them and reconstruct them from alternative foundations. In the domain of writing, AI-driven conversational agents are being used as collaborative writing assistants that help plan, compose, and refine text through dialogue. These agents go beyond static grammar checkers or auto-completion; they engage in an iterative process with the writer. For example, VISAR is an AI-enabled argumentative writing assistant that demonstrates many of these capabilities (Zhang, Gao, Dhaliwal & Li, 2023). VISAR allows writers to brainstorm and organize ideas by creating a visual outline of claims, sub-claims, and evidence, synchronized with a text editor. The agent can generate initial drafts or summaries for each part of the outline, providing a starting point that the user can then edit. Crucially, VISAR supports refining arguments: it offers “argumentation spark recommendations” to improve persuasiveness, such as suggesting evidence to back a point or highlighting a potential counterargument the author should address. In effect, the system helps writers both develop their arguments further and anticipate objections, maintaining a logical flow in the hierarchy of their essay. Another line of research explores conversational agents as discursive thinking partners during writing or design. One such agent acts as a "dissenter" in group design discussions, deliberately challenging ideas to spur critical reflection (Lee, Hwang & Lee, 2024). This LLM-based agent listens to the group’s concepts or a user’s draft content and then poses counterarguments or tough questions as a human critic might. By opposing prevailing opinions or pointing out weaknesses in arguments, the agent encourages users to rethink and strengthen their positions. Such an agent illustrates how AI can raise natural audience questions and counterpoints: for instance, if a student writes an essay advocating a policy, the agent might ask "What about potential downsides or exceptions to your proposal?" and even suggest alternative narratives or examples that contradict the writer’s stance. Engaging with these challenges helps the author deconstruct and expand their arguments, ultimately leading to a more robust and nuanced piece of writing. Recent work explores the extent to which LLMs exhibit a form of theory of mind (ToM) – the capacity to attribute beliefs, desires, and intentions to others – and how this can be applied in writing support. As shown in recent evaluations, modern LLMs can pass classic false-belief tests and demonstrate reasoning about others' knowledge states in stylized settings (Kosinski, 2023; Ullman, Liang, Kosinski & Tenenbaum, 2023). These ToM capabilities enable a writing agent to anticipate how different audiences might interpret or challenge a piece of text, to personalize explanations based on assumed reader beliefs, and to generate content that aligns with audience expectations or queries. For instance, a writer might ask the agent, "Would a skeptical policymaker buy this argument?" and receive a targeted critique based on likely interpretive stances. Evidence from the social cognition literature suggests that effective use of ToM in AI contributes to maintaining socially relevant, pragmatic dialogue, where the agent attends to shared knowledge and informational gaps (Dunbar, Healey & McCabe, 2022).This links closely to efforts in evaluating conversational implicature, discourse relevance, and discourse ethics. Writing agents can be prompted to assess whether a given response respects principles of cooperative dialogue (e.g., Gricean maxims), whether it addresses the question meaningfully, and whether it upholds discursive norms such as relevance, honesty, and justification. Agents can thus simulate feedback such as: "This point evades the question" or "The example violates the presuppositions of the original claim," helping writers maintain ethical and communicatively coherent discourse. These functionalities, based on interactive reasoning about intentions and audience expectations, make modern agents more than text generators: they become responsive interlocutors capable of guiding users toward better arguments, ethically framed claims, and richer engagement with alternative views. Conversational writing assistants can function as real-time editors and tutors. When a user provides a rough draft or an outline, the agent can generate a more polished summary or fill in sections while respecting the context and style specified. The user might then say, "I'm targeting a general audience—simplify this explanation," and the agent will adjust the text accordingly. Likewise, if the user is stuck developing a point, they can ask the agent for an example or a different perspective, effectively getting instant feedback and ideas. By iteratively refining content through prompts and critiques, human writers remain in creative control but benefit from the agent's suggestions for clarity, coherence, and depth. The examples from VISAR and the critical-thinking CA show that modern AI can engage in a back-and-forth writing process – helping to outline content, improve arguments, counteract one-sided reasoning, and ensure the final narrative is well-rounded.

*Analysis→Interactive production of plausible inferences*: interactive dialogical agents can be prompted and contextualized to shape and filter data, to suggest methods for extracting patterns from it, to implement algorithms that process data so as to do so, to refine and augment the space of plausible inferences, and to engage in inductive, deductive and abductive inferences about the implications of the inferences to a predicament or situation. LLA's are increasingly being applied to support analytical reasoning and data exploration tasks. Rather than manually applying statistical methods, users can collaborate with AI agents that guide them in shaping data, selecting appropriate analysis methods, and drawing inferences. StatZ is one such agent, enabling users to perform statistical analyses via natural language conversations (Wen, Kochhar, Zeyada, Javaheri & Rawassizadeh, 2025). A student might ask, "How do I test if this treatment effect is significant?" and the agent will walk through choosing the right statistical test and execute it. StatZ significantly outperformed traditional statistical software in terms of accuracy and efficiency, while reducing cognitive load and supporting intuitive learning. More complex agents like Data Interpreter demonstrate how LLMs can execute full analytical workflows. This system decomposes tasks into sub-tasks and iteratively refines its own methods and outputs (Hong, Lin, Liu et al., 2024). For example, it might begin with an abductive guess about interesting patterns in the data, use clustering or regression to test it (inductive reasoning), and refine the scope of hypotheses based on findings. By maintaining an evolving map of plausible inferences, such agents support an end-to-end analytic process from problem framing to insight generation. These systems also excel in dialogue-driven reasoning. A researcher might hypothesize, "X is related to Y," and the agent will test it, explain results, and surface related patterns: "Actually, Z seems more predictive." This cycle supports inductive learning (from pattern to principle), deductive validation (from theory to data), and abductive innovation (from data to hypothesis). With their knowledge of analytic techniques and real-time feedback, conversational agents become partners in dynamic, multi-stage reasoning.

*Coding →Interactive production of reliable, useful, machine-executable code*: interactive co-pilots and artificial ‘pair programmers’ can be designed and used to synthesize algorithms that solve a specific problem (or to formulate the problem at a sufficiently precise level to admit of a algorithmic solutions), to produce sample proof of concept code that implements an algorithm, to provide feedback on code already built, to generate code on the basis of pseudo algorithmic specifications of tasks and data structures, to generate test vectors for pseudo algorithms and existing code bases, and to integrate and coordinate different algorithms at the level of their interfaces and data utilization. Large language models have made it feasible to have an AI pair programmer that collaborates through natural language. These agents can translate a plain-text task description into functional code and refine it through interaction. AgentCoder exemplifies this multi-agent approach, integrating a “programmer” agent with testing modules (Huang, Bu, Zhang, Luck & Cui, 2023). One agent writes the code, another generates tests, and a third executes them and returns errors. The loop continues until the program passes the tests, resulting in robust, validated implementations. These systems also support educational workflows. For instance, novice coders can describe a problem, get sample code, and then ask follow-up questions about in-place operations or time complexity. LLM-based agents adjust responses accordingly, providing tailored support and debugging advice (Akhoroz & Yildirim, 2025). Unlike static tutorials, these agents enable real-time clarification, customization, and coaching. Agents are adept at converting pseudo-code into executable code and at coordinating interdependent modules. A user might sketch an algorithm in comments or high-level steps, and the agent will build working code with proper data structures. Moreover, conversational agents can generate unit tests, edge case checks, and usage examples, ensuring the code is not only correct but also comprehensive. Research on iterative feedback loops has shown that even modest LLMs can outperform one-shot methods when allowed to interact with user hints and test results, underscoring the value of dialogue in programming.Together, these capabilities make LLM-based coding agents effective interactive programming assistants that support both code creation and conceptual understanding.

In each case, tasks that were previously exercised individually, (read a report, make notes, produce a summary that specifies the key assertions, arguments and grounds of the document) are now exercised in dialogue and interaction with machines (*prompt* an LLM to produce an abstractive summary focused on…, ask an LLM agent to answer a set of questions about the report for the purpose of…. to an audience comprising….; verify its answers with a set of people or databases, design an LLM agent to articulate the key substantive differences between this report and some other vis a vis a background set of assumptions, check for validity on the resulting document by quickly, clearly and incisively querying *people* or machines …). Acts of learning or engagement in knowledge generation, validation or extension processes have critical epistemological (how do you know?) and ontological (what counts as real?) dimensions and consequences that highlight the links between core skills to metacognitive skills. Not surprisingly, the base of skills and activities that enable and comprise the *interactive* consumption and production of text - including text embodied in audio and video signals – and other media (like, generated videos, audios and images) are different enough from their individually exercised counterparts that we have good reason to want to identify, measure and develop these skills independently, and to re-design current learning practices to help us do just that. And, as it turns out, even in the case of individualistic tasks that have already become heavily 'socialized' during the past two decades (like coding/programming →interactive compilation/paired programming), the use of LLA's as the 'intelligent other' induces a need for a set of skills that only overlap those required when interacting with human collaborators, but extend beyond them [Sarkar et al, 2022].

Key to such a re-design is the realization that the basic palette of operators and tasks whose competent performance is the hallmark of the 'skilled individual' –was to draw upon in more individualistic eras and times - such as *registering and recalling on cue*, summarizing - extractively or abstractively - making deductive, inductive or abductive inferences on the basis of textual or symbolic structures that encode models, narratives and other 'knowledge structures' - are now being augmented and sometimes replaced by a new set of *interactive* tasks and operations. The individualistic tasks associated with the intentional and sophisticated consumption and production of text (reading and writing, in both natural and artificial (computer) languages, in written and spoken modalities) are now replaced and augmented by tasks that have to do with the production of *communicative acts* in a context, for a purpose, with a pre-text and a sub-text.

For instance:

'*Reading an essay about X'* becomes not only:

> parsing the semantics and syntax of the text with a view to recovering the substance, relevance, novelty and implications or applications of the text to a predicament, situation or topic –

but also:

> articulating a set of queries and questions that one hopes to have answered as a result of having read *X for a purpose* - and -
>
> duly contextualizing and prompting a GenAI agent to parse the text through the prism of these questions – or, indeed –
>
> formulating specific questions, queries and challenges on the basis of generic patterns of successful inquiry.

*'Writing a report about X'* is not only putting together a series of communicative acts or utterances such as:

> articulating and expressing *a set of claims and positions that are warrantedly assertible to a person of a certain disposition and background, with certain interests and affordances for the purpose of informing, persuading…*
>
> *explaining a fact, principle, model, method or other generality to someone who has particular standards for determining whether or not and how an explanandum explains an an explanans,*
>
> *justifying a maxim or an action or a decision to someone who has specific pre-dispositions or commitments, etc.*

but also:

> *clarifying* both the referents of terms and the truth values dependencies and consequences of assertions in answer to queries and questions produced in real time by a GenAI agent;
>
> prompting the agent to raise such questions in a way that mimics the conversational dynamics of dialogues one expects to have;
>
> *answering* questions about validity, informativeness, relevance and the intent of the writer or speaker in making assertions;
>
> *responding* to challenges raised to the assertive force of an expression, the validity, completeness and relevance of a statement and the coherence and consistency of an argument.

The monological and individualistic processes of the *reading-writing* nexus are replaced by dialogical processes of *reading-alongside* and *writing-with*. Accordingly, the repertoire of communicative acts that need to be mastered range beyond asserting, explaining, justifying or in some way advocating, and include *interrogative* skills (querying, questioning, verifying, challenging in ways that jointly heed the objectives of increasing clarity, validity, coherence and transparency and inducing the respondent to answer so as to do so as well) and *responsive* skills (answering and responding in informative and relevant ways). Taken as a whole, these moves become powerful pedagogical tools that escalate the cognitive involvement of humans to a meta-cognitive level. In the examples above, routine, mundane, and individual learning activities are managed through AI and more complex and involved tasks are initiated by humans through more thoughtful prompting and assessment.

*But Was not Cognition Always-Already Interactional?*

To those familiar with a certain tradition of thinking about language and cognition [Vygotsky, 1981; Sperber and Mercier, 2011; Mercier and Sperber, 2012; Wittgenstein, 1958; Bloor, 1983], the 'interactional turn' in the definition and cultivation of intelligence will not be as surprising as to others. In this tradition, thinking is a form of internal conversation [Vygotsky, 1981] which, in turn, is an internalized version of the sorts of dialogical exchanges a person has [Moldoveanu and Martin, 2009]. Because reasoning is dialogical and dialectical in nature, measuring one's skill and prowess therein needs to account for the communicative – justificatory, explanatory, clarificatory – functions that reason plays in society [Sperber and Mercier, 2011; 2012]. Many of the 'enigmas of reason' – including *non sequiturs* and *incorrigible fallacies* [Johnson Laird, 2007] unwarranted biases and apparently sub-optimal heuristics [Tversky and Kahneman, 1979] appear as reasonable ecological adaptations of communicative practices to interactional predicaments, such as having to *persuade* or convince someone of the validity of a claim or having to figure out if one's interlocutor is telling the truth. These practices are far more textured and subtle than the 'fast and frugal heuristics' [Gigerenzer et al, 2011] that have been proposed to explain the purely cognitive engagement of human minds with large information sets under time pressure and resource constraints: they are at least as sophisticated and universal as the 'deep learning' networks that are universal function approximators [Goodfellow and Bengio, 2017], as they need to adapt not only to 'models', 'data' and 'objective functions', but also to the interpersonal, social, cultural, physical and physiological context of the communication based on the models and the data.

The ‘interactionalist’ turn in learning and knowledge building processes that we propose here has deep roots. To date, however, it has not been fully articulated and developed into a set of models, methods and blueprints for seeing, measuring and developing skills and for the design and development of learning experiences: tests of intelligence, cognitive and algorithmic skill and computational prowess remain focused on the efforts of an individual-in-isolation. While non-cognitive measures of individual skill have been advanced in the psychometric and econometric literatures [Borghans and Heckman, 2008], the tasks or (self-reported) traits they are based on remain individualistic in nature, as in the case of personality traits (‘reliable dispositions to act in certain ways across a range of different situations’). By contrast, the ‘interactionalist’ approach to human skill specification, learning design is significantly under-theorized and under-developed: We do not have good interactional and dialogical skill ontologies – let alone measures we can reliably gauge learner progress against. Developing such an ontology – and associated measures and means for developing these skills – present a significant opportunity to advance both human skill classification and human up-skilling practices to the stage they need to arrive at in an era in which co-sapience becomes widespread.

*Dialogical Learning for Interactional Skill development:*

*An Architectural Solution* for *Higher Education*

Suppose you wanted to learn how to play tennis as quickly and proficiently as possible. There are lots of tennis training programs and academies, and all of them are based on the same philosophy of tennis learning: you have to become a good short, medium and long distance runner first, then you have to specifically learn how to jump forward, backward and sideways, then go through a set of weight training classes to develop your anterior deltoids and biceps of the swinging arm, then master walking on a balance beam, then you will get to swing a racket while looking at yourself in a mirror, and you will graduate after you keep up a ‘rally’ against a fixed wall at least 100 times in a row without a glitch. Once you graduate, you will get to play against real people, wielding rackets and trying to win points from you by doing all people do when they try for real to win. This absurd picture of a ‘tennis education’ is not a far-fetched model of the current divergence between what most of higher education models teach and help learners learn, on one hand, and the sorts of tasks learners are expected to be able to perform competently by their recruiters and employers after they graduate.

Interactionalism posits matters *need not be this way*, and indeed, that they cannot *remain* this way if higher learning is to deliver on its educational and upskilling mission and social contract to provide a vehicle for socio economic mobility through the acquisition of valuable skills. Rather, it envisions a re-design of the experience of all learners so that *the method of learning embodies the desired learning outcome*. Instead of requiring learners to submit to schedules of information consumption and a canon of tests and assignments that test the competent exercise of individually exercised skills to perform tasks such as recall and interpretation of information, modeling of phenomena and inferences of patterns and regularities from the data sets these phenomena have generated, the articulation of plausible arguments about - and exegeses and critiques of - textual forms of information, and the calculation of solutions to well-specified problems using deductive, inductive and abductive patterns of inference, the interactionalist program *makes the learning method itself a model for the new, interactional skill set learners will require*: "The learning experience embodies the learning objectives ". It shifts the emphasis of learning from an individual production of artifacts to an interactive production of artifacts.

Crucially, interactionalism also shifts the *basis for evaluation* from the competent production of an artifact (an essay, a paragraph, a piece of code, an algorithm, a proof, a computation for optimizing, predicting, sorting, searching, etc) to a *pattern of interaction* with an appropriately patterned dialogical agent, aimed at co-producing an artifact. For instance:

> *From individual calculation to interactive computation:* replace (or augment) the step-by-step record of the calculation and the final result as the basis of evaluation, by a transcript of the interaction with the dialogical agent aimed at *producing the fastest way of producing a reliable computational procedure* for solving a problem or providing the proof of a conjecture or theorem – the embodied aim of algorithmic analysis and optimization;

*From individual modeling to interactive model construction, elaboration and testing:* replace the articulation of a model (a set of independent and dependent variables, co-variates, modulators and moderators, relationships among them and boundary and initial conditions) of a trend, fact base, or 'phenomenon' as the basis for evaluation by a transcript of a record of interaction with a dialogical agent aimed at producing a model *for* optimizing, predicting, or explaining a set of data, which heeds a set of logical, material or computational constraints and includes the formulation of queries, questions, challenges and prompts, and the interactive specification of critical tests of the validity and reliability of the model;

*From individual argument production to dialogical argument construction and refinement:* replace the syllogistic forms of valid and sound arguments for explaining, justifying or deconstructing particular or general statements as the basis for evaluation with a transcript of an interaction with a learning agent aimed at a dia-logical construction of an argument that satisfies a set of conditions such as intelligibility, coherence, or logical consistency with other assertions or arguments and which is robust to various challenges to its grounds and inferential links;

*From individual exegesis to interactive interpretation:* replace the individually produced essay or paragraph that lays out plausible, warranted interpretations of a piece of text vis a vis a set of pre-existing substantive, epistemological or ontological commitments and a set of intentions of the writer with a transcript of an interaction with a dialogical agent aimed at reconstructing plausible interpretations and adducing evidence for their validity from the target text, refining the interpretations in view of evidence from the target text or other texts, and integrating across different plausible interpretations;

*From individual to interactive verification:* replace the individually produced report of the verification of the validity, soundness, completeness, reliability, generalizability, etc. of a piece of text with a transcript of an interaction with a dialogical agent that aims to produce a verification protocol (a set of queries, questions, challenges, putative answers) whose outputs would provide sound reasons for acting upon the basis of the predictions and results of a piece of text.

*The Metacognitive and Meta-Emotional Demands of Interactionalism as Developmental Opportunities for Learners*

Interactionalism matches the 'phase change' in the nature, exercise and deployment of human skills that GenAI enables in the workplace by augmenting the dimensionality of 'knowledge work' to include a meta-cognitive and a meta-emotional dimension. Prompting, patterning, and contextualizing LLA's, designing tests for these agents that sample across the possible worlds (scenarios, problem instances, use cases, user types) in which are instantiated, and architecting collections of agents that perform a specific workflow – which includes the tasks of *interacting* with other people and machines - requires a form of thinking that is different *in kind* from that required to interact with machines whose task execution is deterministic and the range of inputs is restricted to formal scripts written in pre-set syntax ('code'). When we engage in such tasks, we are monitoring, designing, patterning and offloading cognitive work in a complex, non-stationary environment [Tankelevitch et al, 2024]. Our ability to estimate the relative strengths and weaknesses of human and AI agents at performing certain tasks, monitor the degree to which AI output on a per-task basis is useful, valid, attuned, responsive, informative, etc, and differentially energize specific cognitive functions (calculative, interpretive, predictive, communicative, etc) while minimizing task switching costs is very important to the quality of both the process and the product of the man-machine interaction. Meta-cognitive activities refer to action sequences aimed at monitoring and controlling one's own mental states and events (thoughts and perceptions) [Stuss, 2011]. Meta-cognitive skills are specific combinations of general abilities to register, monitor and purposefully control one's own mental events in ways that are useful and adaptive to the environment and to the problem or predicament (see Table 1). They include task planning, prospecting, partitioning and sequential energization, task switching and attention sharing, monitoring of one's own mental events and sensory stimuli, and the suppression of impulses to subordinate them to the pursuit of a coherent purpose.

| **Meta-Cognitive Task** | **Sample Instance** | **Meta Cognitive Skill Class** |
|---|---|---|
| Self Explicitation | Specifying one's objective and constraints, in advance of solving a problem or making a decision | Monitoring, Registering, Control |
| Task Specification | Specifying *how* one will frame and solve a problem in advance of solving it | Monitoring, Control |
| Task Decomposition | Breaking down a task into spatio-temporally contiguous and causally continuous sub-tasks | Monitoring, Control |
| Sub-task energization | Allocating energy to sequential tasks in a sequence, suppressing impulses that take one off-task. | Monitoring, Control |
| Task Performance Evaluation | Evaluating the outcome and payoff of the performance of a task | Registering, Monitoring, Control |
| Task Switching | Switching among different tasks | Monitoring, Control |
| Partial credit assignment | Assigning partial credit to the components of an implemented task plan, on the basis of registering feedback on task performance | Registering, Monitoring |
| Objective Refinement | Refining one's objective function in response to the outcome of the implementation of a task | Monitoring, Control |
| Task Specification Refinement | Refining the procedural or algorithmic structure of a task in response to feedback on the outcome of the implementation of the task. | Monitoring, Control |

*Table 1. A Novel Decomposition of Meta-Cognitive Skills that Connects General Meta-Cognitive Abilities to Task-Specific Skills, across a broad range of tasks.*

This decomposition of meta-cognitive skills allows us to directly connect them to the specific skills that an interactionalist approach to learning develops. The prototypical and modal ways in which people interact with LLM's require a dense and proactive exercise of meta-cognitive functions. Consider, for instance:

- Designing an LLA instruction set (or, 'prompt') – which requires iterative switching (meta-cognitive) between mapping out the likely *response characteristic* of an LLA to the prompt's specific language (meta-cognitive) to the design of the language of the prompt itself (cognitive);
- Refining an LLA prompt – which requires rapid switching (meta-cognitive) between iterative testing of the response function of the LLM to the prompt and mapping 'misfires' and 'malfunctions' onto the various parts of the prompt (cognitive);
- Specifying the objectives of a (LLA) – which requires quickly switching (meta-cognitive) between user specifying response characteristics to the agent's functions and articulating putative goals or objectives whose pursuit will improve the LLA function across many different types of users (meta-cognitive);
- Specifying the chained tasks of an LLA – which requires rapid alternation (meta-cognitive) among ascertaining likely response patterns across many user types (meta-cognitive) to a first pass at a system prompt and iterative (meta cognitive) modification of the objectives and tasks of the LLA.
- Designing an agentic architecture and workflow that replicates, across a range of environments, the differential response characteristics and intelligent behavior of a human who interacts with that agent (meta-cognitive);
- Designing algorithmic processes and procedures that guide and shape the response of a piece of code produced by an LLA to best-case, worst case and average case inputs (meta-cognitive);
- Specifying evaluation rubrics and metrics for eliciting measurements of the success with which a Large Language Agent was deployed (meta-cognitive);
- Specifying gold standard, average case and substandard responses of a special purpose LLA to an intended user's inputs (meta-cognitive),

which can guide the deployment of rubrics and metrics for overall quality of response evaluations;

- Designing an agentic architecture that iteratively tests for the user response characteristics and dynamics of an LLA, elicits precise human feedback on pieces of an interaction, and proposed changes to the design of the target LLA that respond to the results of the evaluation;
- …

These cases highlight not only the interactive and collaborative aspect of the human-AI interface, but also the universality and accessibility of the interaction and design language ('natural language', for the most part) that designers use to interface to LLM's via LLA's. The accessibility of the operating system – via prompts and contextualizations that are free of the syntactic and semantic idiosyncrasies of programming languages – democratizes access to the human-computer interface and enables the development of a wide range of skills for anyone who can take a meta-linguistic approach to the design of his or her own communicative acts.

These concrete tasks can be mapped into the task-wise decomposition of meta-cognitive skills to provide a 'spectrogram' of the skills required to design and pattern LLA's:

| Meta-Cognitive Task | Sample instance from Large Language Agent Design |
|---|---|
| Self Explicitation | Specifying the objective function or purpose for an LLM-based agent (an LLA) |
| Task Specification | Specifying the algorithmic, heuristic or procedural structure of a task that an LLA is designed to perform |
| Task Decomposition | Specifying a decomposition of tasks to be performed by one or more LLA's |
| Sub-task energization | Designing an LLA on the basis of alternating shifts between persona, objective and procedural specifications; alternating between the production of normative/counternormative 'sample responses' and the specification of procedural details |
| Task Performance Evaluation | Designing evaluation metrics for the interaction between an LLA and an end user |
| Task Switching | Alternating, iteratively and quickly, between objective, constraint, persona, and task structure specifications in an LLA design process. |
| Partial credit assignment | Assigning credit and blame for LLA performance to the different components of the LLA blueprint (objective, persona, context, procedure, etc) |
| Objective Refinement | Refining the objective of the LLA in response to performance feedback |
| Task Specification Refinement | Refining the specification of a task for an LLA in response to performance feedback. |

*Table 2. Mapping of meta-cognitive tasks requiring meta-cognitive skills to the tasks interactionalist models of learning and work entail.*

In addition to the meta-cognitive dimension of competence, designing, testing, evaluating, deploying and refining LLA's also taps into a meta-emotional and meta-relational dimension of human capabilities. Simply put, if LLA's are to function as extensions of one's working self and if one's work is densely interactional and interactive, then the design of such agents must consider the ways of being of all those humans who will in turn interact with this augmented self.

Such tasks as:

- Making accurate, reliable inferences about the emotional response of the end user of a Large Language Agent;
- Creating robust measures of the emotional and relational connectedness such a user might feel to the agent;
- Creating realistic instances of a 'prototypical user' that can function as a tester or a provider of test cases –

- Testing the ‘Theory of Mind’ capabilities of a large Language Agent in complex, dialogical use cases –

all make use of a meta-relational and meta-emotional skill set that current educational systems struggle to develop, given the impoverished interactional landscape of the lecture-based classroom [Moldoveanu and Narayandas, 2022; Moldoveanu, 2024].

| **Type of Meta-Skill** | **Instance of Interactionalist Task Making use of the Skill** |
|---|---|
| Meta-Emotional: infer intentionality | Make accurate, reliable inferences about the intentionality a user of an LLA infers from her interactions with it |
| Meta-Emotional: infer emotionality | Make accurate, reliable inferences about the emotional state that an LLA response induces in its user |
| Meta-Emotional: measure emotionality | Design evaluative rubrics and measures for the emotional connectedness and attunement; design tests and prototypical ‘tester personalities’ for such. |
| Meta-Relational: interactive reasoning | Design agents that account for what the user thinks and what s/he infers the agent ‘thinks’ s/he thinks to achieve higher levels of intimacy and connectedness |
| Meta-Linguistic/Meta-relational | Design agents that heed the specific style and turns corresponding to the persuasive use of language in given socio-economic setting. |

*Table 3. Dialing up Meta-Skills to the Meta-Emotional and Meta-Relational Realms*

Far from introducing a need to 'simplify' or 'dumb down' the process by which humans interact with Large Language Models, the new learning landscape enabled by the need for Large Language Agents and people that can pattern, create, deploy, test and refine them reveals a new frontier of 'meta-human' skills that educational systems need to take upon the themselves to help learners of all ages develop. While a retreat to the 'comfortable truths' of designing human-computer interactions for the information age (eg 'keep it simple & make it happen) is tempting, there are good arguments to resist the simplification itch in this case [Sarkar, 2023]. The interface between humans and machines is no longer smooth and stable, but, rather, jagged and volatile [Dell'Acqua et al, 2023]. Sometimes the LLA designer needs to specify tasks at a mechanical 'base layer' to successfully automate higher level 'reasoning' across a wide range of contexts. Sometimes the contexts of LLA use change quickly enough that all the reliability/validity guarantees of the 'probably approximately correct' learning framework that undergirds all of machine learning need to be discarded and new data sets need to be created. And, LLM's absorb and amplify the 'tips and tricks' of successful LLA designers and users. One size will not fit all – or even most: human cognition needs help to rise above itself, and educators who take their roles seriously need to internalize their new tasks.

*Building Blocks for a Meaningful GenAI-Driven Re-Engineering of Higher Education*

Interactionalism introduces an important 'meta-human' dimension to the ontology of skills a higher education system can seek to select for and cultivate – one that, as the name indicates, sits in a 'meta' position vis a vis the usual set (motor, perceptual, cognitive, emotional, relational) ones. Traditional measures of ability – such as working memory, processing speed, verbal and textual comprehension, pattern recognition and pattern matching – alongside more recently recognized measures of social and emotional intelligence – need to be augmented by skills that specifically address the meta-cognitive, meta-emotional and meta-social dimensions that working with, through and alongside GenAI agents require. And, because of the speed with which human collaboration with human designed AI agents is diffusing through institutional and organizational environments, there is a time-sensitive need to re-design the learner selection, experience and evaluation methods to adapt to the new skills requirements of labor markets.

*Learner Selection.*

It is an time-tested adage that 'we get what we select for' – conditional upon the accuracy and reliability of the instruments we use for selection. Current selection instruments for higher education environments – ranging from standardized tests of cognitive and technical skills to tests of written and oral comprehension to tests of background disciplinary knowledge to admission interviews that are carried out by often untrained human evaluators and feature one-shot answers to more or less standard questions can and should be re-designed for the interactionalist environment. The direction of the change vector is provided by the meta-cognitive and meta-emotional dimensions of human skill we wish to develop. The instruments of the change are provided by LLAs themselves. It is now both possible and desirable to move from 'one shot' assessment vehicles – such as multiple choice and short answer questions, calculations, pieces of code and pseudo-code and one-shot oral answer questions – to multiple-shot evaluations – carried out under timed, secured and invigilated environments – in which the candidate needs to answer several, chained questions, challenges and invitations to elaborate. These questions can test not only *what* the candidate knows and *what* s/he can do with what s/he knows in a limited number of test cases – but also how she deals with questions that probe into *how* she knows what she knows, *what* the epistemological and methodological bases for her assertions are, *how* she would verify or falsify the validity, reliability, and generativity of various artifacts (claims, arguments, algorithms, models, theories) and *what* the range and span of queries, applications, and extensions of a knowledge base to an unknown or unfamiliar field of knowledge might be. And multimodal interactive admission agents can be patterned to elicit timed oral responses to challenging questions and provocative statements that test for a candidate's non-verbal and often non-cognitive abilities to evince presence, competence and trustworthiness [Moldoveanu, 2024]. Of course, the use of dedicated custom agents is but a transitional step: The specific empowerment and enablement of current staff and faculty members engaged in the all-important process of selecting students in the design, prototyping, evaluation, refinement and deployment of agents that embody the specific intuitions and values of each institution – that will allow both educational product differentiation and continued progress on selection processes.

*Learning Experience and Environment.*

Large Language Agents enable a re-design of the learning experience and environment of the learner within curricular programs that privilege the specific competencies and skills we expect humans as wielders of GenAI tools to be able to show proficiency in. Clearly, many of the canonical tasks (such as data analysis and interpretation, model formulation and specification, interpretive analysis and framing of complex text) that cut across natural and social sciences and humanities programs can be re-designed and refashioned for an interactionalist environment, in which LLA's, duly designed, take on a large component of the task. This leaves the precise articulation of the meta-human skill set, alongside discipline-specific examples of 'use cases' in which these skills are exercised, a key component of the re-design of the learning environment.

A second, equally important component of the re-design of the learner's experience is the introduction of a radically dialogical approach to learning itself, via a dialogical grammar that captures the key moments (acceptance of the learning scenario, personalization of the conceptual content to the interest of the learner, contextualization of assessment to the specific skills and capabilities we seek to have the learner develop) of the learning process. Well-designed LLAs can be used to provide 'always-on' feedback to learners on artifacts produced on demand. The 'broadcasting' model of learning, featuring a high bandwidth 'downlink' (teacher→learner) channel, a low bandwidth (learner-teacher) channel (comprising assignments and quizzes and class presentations) and a very low bandwidth feedback channel (teacher→learner) featuring feedback on learner-produced artifacts ('assignments') can now be effectively replaced by an 'always-on' instruction-query-answer-feedback loop that massively accelerates the learning of any particular skill and the range of situations and instances to which the learned competency is applied.

In both cases, the key to broad adoption and deployment is the empowerment of instructors to design agents, agentic frameworks, and dialogical scenarios that represent their own expertise and pedagogical vision. This will require a rapid evolution of the industry from the use of dedicated, specialized solutions – such as virtual teaching assistants and graders that embody a specific language system, dialogical structure and pedagogical approach, to the enablement of *user innovation* – from instructors and educators keen to take their work into the LLA ecosystem.

*Learner Evaluation*

By now, it would have become clear that learner evaluation can and should *also* become an always-on, continuous process. Attuned, astute tutors can evaluate learners based on the questions, challenges and other behaviors learners produce during the learning process itself – as well as on the basis of the answers and other artifacts learners give upon being prompted to do so by tests, assignments and exams. There is no distinction between 'online' and offline' behaviors – just as there are none in the workplace or a tightly connected, attuned seminar setting in which every participant has a stake in the discussion. If we have – as we now do - the ability to continuously monitor and evaluate learner behaviors during the learning process, then incisive, relevant questions and interjections, cutting and relevant challenges to the 'teaching agent and responses to productively ampliative invitations to elaborate on a point can be as informative as assessments of learner performance on tests and quizzes. We get, into the bargain, a heightened level of learner attunement, agency and participation into the instructional process itself, which can come to approximate the dynamic of the exemplary 'tutor/mentor' relationship that is at the core of Bloom's [1984] insight. becomes guided by the learner's own questions and challenges, and personalized by the learner's own prior background knowledge and experiences. And, as before, the active participation by instructors and staff – by designing agents that assess, evaluate and provide feedback on student work – will be critical to both the widespread adoption of LLA's in the evaluation process, and in advancing evaluation of students into the dialogical era.

***

Interactionalism thus 'eats its own cooking': it makes use of the very skill set it is aiming to help learners develop as a blueprint for the re-design of the learner selection, experience, environment and evaluation process. It highlights the importance of a 'phase shift' in the construal of LLM uses in education – from 'what AI can do' to 'what we can do with AI' – and specifically to 'what we can do with AI to help learners do more with AI'.

*Conclusion*

We have articulated the elements of an 'interactionalist turn' and associated agenda in higher education, enabled and facilitated by the broad deployment of Large Language Agents and Large Language Models. Key to its successful development and deployment at scale is a shift in speaking and thinking about GenAI technologies in terms of what they can do to one where we speak about *what we can do with them* through the design of Large Language Agents that use LLM's as 'operating systems'. We highlighted the importance of 'meta-human' skills in the successful augmentation, replication and automation of human workflows and tasks using Large Language Agents, and argued that current emphases on *reducing* meta-cognitive workloads required for humans to interact with Large Language Models should be construed as an opportunity for learners to develop an important and sophisticated – but nonetheless articulable and measurable – skill set that will enable them to be *creators and co-creators* rather than users and consumers of Large Language Agents and their outputs.